\documentclass[12pt]{article}

\usepackage{epsfig,graphicx,bm,amsmath,amsfonts}
\usepackage{epstopdf}

\newcommand{\ben}{\begin{eqnarray}\displaystyle}
\newcommand{\een}{\end{eqnarray}}
\newcommand{\be}{\begin{equation}}
\newcommand{\ee}{\end{equation}}
\newcommand{\lb}{\left (}
\newcommand{\rb}{\right )}
\newcommand{\ltb}{\left [}
\newcommand{\rtb}{\right ]}
\newcommand{\nn}{\nonumber}

\newcommand{\bea}{\begin{eqnarray}}
\newcommand{\eea}{\end{eqnarray}}
\def\be{\begin{equation}}
\def\ee{\end{equation}}
\def\ba{\begin{eqnarray}}
\def\ea{\end{eqnarray}}
\def\nn{\nonumber \\}

\begin{document}

\setlength{\unitlength}{1mm}

\begin{titlepage}

\begin{flushright}
AEI-2011-022
\end{flushright}
\vspace{2cm}

\begin{center}
{\bf \Large Near horizon data and physical charges of extremal AdS black holes}\\
\vspace*{1cm}
\end{center}

\begin{center}
\bf{Dumitru Astefanesei$^{a}$,}
\bf{Nabamita Banerjee$^{b}$,}
\bf{Suvankar Dutta$^{c}$}

\vspace{.5cm}

{\small \it $^{a}$ Max-Planck-Institut f\"ur Gravitationsphysik, Albert-Einstein-Institut, 14476 Golm, Germany}\\
\vspace{2mm}
{\small \it $^{b}$ITF, Utrecht University, Utrecht, The Netherlands}\\
\vspace{2mm}
{\small \it $^{c}$ Department of Physics, Swansea University,
Swansea, U.K.}\\[.5em]

{Email: \small {\tt dumitru@aei.mpg.de, \ N.Banerjee@uu.nl, \ pysd@swan.ac.uk}}

\end{center}

\vspace{1cm}

\begin{abstract}
We compute the physical charges and discuss the properties
of a large class of five-dimensional extremal AdS black
holes by using the near horizon data. Our examples include
baryonic and electromagnetic black branes, as well as
supersymmetric spinning black holes. In the presence of the
gauge Chern-Simons term, the five-dimensional physical charges
are the Page charges. We carry out the near horizon analysis
and compute the four-dimensional charges of the corresponding
black holes by using the entropy function formalism and show 
that they match the Page charges.

\end{abstract}

\vspace{2cm}


\end{titlepage}

\tableofcontents

\section{Introduction}
Black hole physics is a very active area of ongoing research. A statistical
understanding of black hole entropy is one of the most important
and long standing questions in theoretical physics.

In the last years, important progress in understanding the attractor mechanism
and entropy of extremal (non-BPS) black holes  was based on the entropy function
formalism \cite{Sen:2005wa, Sen:2005iz, Astefanesei:2006dd, Sen:2007qy}. An
important advantage of this method is that one
needs just the near horizon geometry to characterize the black hole. In particular,
one can compute the entropy and the conserved charges by using just the
near horizon geometry \cite{Sen:2007qy,Hanaki:2007mb}.

In \cite{Hanaki:2007mb} it was explicitly shown how to construct the conserved
charges of black holes/rings by using just the near horizon data. For example,
it is by now well known that the dipole charge appears in the first
law of black ring thermodynamics, even if it is not conserved \cite{Copsey:2005se}.
Therefore, the entropy of the black rings can also depend on the dipole
charges, not only on the usual conserved charges.

In general, when there are degrees of freedom living outside the
horizon, they
contribute to the asymptotic charges and so the asymptotic charges
and the charges
computed from the near horizon data may not be the same.\footnote{A
discussion on asymptotic charges and the charges computed at the
horizon for the black ring can be found in \cite{Astefanesei:2009wi}.}
One expects that the macroscopic entropy
is completely determined by the near horizon geometry. However, a counterexample
is the $4D$-$5D$ lift of BMPV black hole \cite{Breckenridge:1996is}. That is, the $4D$ and $5D$ black holes
have the same near horizon geometry, but different microscopic
spectra. If true, that would imply that different microscopic
entropies will correspond to the same near horizon geometry.
In fact, this discrepancy \cite{Banerjee:2008ag,Castro:2008ys} was solved in \cite{Banerjee:2009uk} by removing the `hair', which
contributes to the degeneracies. This further suggests that the
physical charges can
be  obtained, in fact,  from the near horizon data only.

Also, in the presence of the gauge Chern-Simons term, there are
subtleties in the
definition of the charges.  Since the Maxwell charge is carried
by the gauge fields themselves, it is diffused throughout the bulk
and so it is not localized. This clearly resembles the previous discussion,
and one expects a `hair' contribution to the asymptotic charges. The physical
charges in this case are the so called `Page charges' (see \cite{Marolf:2000cb}
for a nice brief review).

A similar analysis can be used for asymptotically AdS black holes. However,
the interpretation of the attractor mechanism is quite different  than
in
flat space. That is, the moduli flow is in fact an RG flow towards the
IR
attractor horizon once the theory is embedded in string theory (more precisely
in type IIB) \cite{Astefanesei:2007vh}. Interestingly enough, the near horizon
data can be also used to compute the shear viscosity coefficient, not just
the entropy \cite{Iqbal:2008by,Banerjee:2009wg,Banerjee:2009ju,Astefanesei:2010dk}.\footnote{For
other transport coefficients
there is, in general, a non-trivial flow and the near horizon data is
not enough \cite{Banerjee:2010zd}.}

Given these motivations, it is clearly useful to understand the robustness
of the near horizon analysis for the AdS black holes. In this paper we generalize
the work of \cite{Hanaki:2007mb} to AdS
black holes. Though, an important difference is that we use the entropy
function formalism to obtain the four dimensional physical charges and
compare them with the charges of \cite{Hanaki:2007mb}.
We would like to emphasize that our analysis is built on the
previous (but less general) work on asymptotically flat and AdS
extremal black holes \cite{Hanaki:2007mb, Astefanesei:2009sh, Cardoso:2007rg, dewit-katmadas}. However, the
near horizon geometry
ansatz we consider is more general due to the existence of a magnetic
Kaluza-Klein (KK) part. We work out in detail several examples, which
are relevant in the context of AdS/CFT duality.

The paper is organized as follows: in Section 2 we present a concrete
analysis of the entropy function formalism for a generic near horizon geometry
ansatz when the gauge Chern-Simons term is present. We also show that
the four-dimensional charges obtained from the entropy function after
KK reduction match the five-dimensional Page charges. In Section 3 we
apply the general results of Section 2 for some concrete
examples. Finally, we conclude with a discussion of our results. In
Appendix A we present details of the KK reduction of the Chern-Simons
term. Appendix B contains an analysis of toroidal spinning black branes.

\section{Near horizon geometry and physical charges}

In this section, we present a detailed analysis of the entropy
function formalism for a general class of $5$-dimensional AdS
black holes. We start with an action that contains a Chern-Simons
gauge term and, to compute the physical charges, we KK reduce
to obtain a gauge invariant action in four dimensions. A similar
analysis for asymptotically
AdS and flat black holes was carried out in
\cite{Astefanesei:2009sh, Cardoso:2007rg} --- for other
applications of entropy function
in AdS space, see \cite{Morales:2006gm}. However,
our analysis
is more general since we work with a generic KK field. In this
way, we can study within the same set-up the SUSY AdS {\it spinning}
black holes and, also, black branes and non-supersymmetric
extremal black holes. The latter are important since yield insights into the
physics of strongly interacting systems at non-zero density
\cite{Herzog:2009gd, D'Hoker:2009mm, Astefanesei:2010dk}.

We start with a brief review of \cite{Hanaki:2007mb} and generalize these results
to AdS black holes (to our best knowledge, an analysis for AdS black
holes did not appear before). Then, we compare these results with
the ones obtained from the entropy function formalism.

Why an analysis of physical properties from the near horizon geometry is
important? Since, in general, it is not easy to construct the {\it full}
analytic black hole solutions in complicated theories, it is clearly
very advantageous to extract information from the near horizon data only. For
example, one can compute the viscosity/entropy ratio just with the knowledge
of the near horizon geometry \cite{Banerjee:2009ju}.

Another important application is in the case of black holes for which there exist degrees of freedom
living outside the horizon (`hair') --- from the point of view of an
asymptotic observer, the hair can contribute to the macroscopic
degeneracy and/or the asymptotic charges. However, the near horizon geometry
should encode the `statistical' information for computing the entropy
(to compare the macroscopic degeneracy with the microscopic degeneracy,
one has to subtract the hair, \cite{Banerjee:2009uk}). Therefore, a computation of
the physical charges at the horizon is important.

Due to the presence of the gauge Chern-Simons term in the $5$-dimensional theory,
the usual `Maxwell charge' is not localized --- this charge is carried
by the gauge fields themselves and so is diffused throughout the bulk. An
important question then is what is the physical charge in the presence
of Chern-Simons term? In \cite{Hanaki:2007mb} it was argued that, in fact, the so called
`Page charge' is the correct $5$D charge. The Maxwell charges in $4$D are obtained
by KK reduction of the Page charges. We will use the entropy function
formalism to show that, indeed, the $4$D charges of \cite{Hanaki:2007mb} match the
physical charges obtained from the entropy function formalism.

The theory we shall be considering is minimal $D=5$ {\it gauged}
supergravity with bosonic action
\bea
\label{gauged}
 S_5 &=& \frac{1}{4\pi G_5}\int \left[\left( \frac{R_5}{4} + \frac{3}{\ell^2}
 \right) \star 1 - \frac{1}{2} F \wedge \star F - \frac{2}{3\sqrt{3}} F
 \wedge F \wedge A \right]\nn
&=&{\frac{1}{(k_5)^2}}\int d^5x \left[\sqrt{-g}(R_5 -F^2+\frac{12}{\ell^2}) -
{\frac{2}{3\sqrt{3}}}\varepsilon^{\alpha\beta\gamma\tau\delta}
A_\alpha F_{\beta\gamma}F_{\tau\delta}\right],
\eea
where $R_5$ is the Ricci scalar, ${}F^2\equiv
{}F_{\alpha\beta}{}F^{\alpha\beta}$, and
$F=dA$ is the field strength of the $U(1)$ gauge field. We also use the notation
$(k_D)^2=16\pi G_D$, where $G_D$ is the gravitational constant in $D$ dimensions.
The bosonic equations of motion are
\bea
\label{eqofmot}
{}^5 R_{\alpha\beta}-2{}F_{\alpha\gamma}{}F_\beta{}^\gamma
+\frac{1}{3}g_{\alpha\beta}{}(F^2+{\frac{12}{\ell^2}})&=&0,\nn d*{}F +
\frac{2}{\sqrt{3}} {}F \wedge {}F&=&0.
\eea

The last equation can be rewritten as
\bea
d\left(*{}F +\frac{2}{\sqrt{3}} {}A \wedge {}F\right)= *j=0
\eea
and so it is this exterior derivative that should be identified
with a current. It is clear that the `Page' current is conserved
and localized (in the usual sense that it vanishes when the equations
of motion hold) --- see \cite{Marolf:2000cb} for a more detailed discussion.

Consequently, we can define a conserved, localized charge as the
integral of the Page current over an arbitrary three-cycle, $\Sigma$
surrounding the black hole:
\bea
\label{page}
(4\pi G_5)Q_{5d}=\int_\Sigma\left(*{}F +\frac{2}{\sqrt{3}}
{}A \wedge {}F\right)=
\int_\infty\left(*{}F +\frac{2}{\sqrt{3}} {}A \wedge {}F\right)=
\int_H\left(*{}F +\frac{2}{\sqrt{3}} {}A \wedge {}F\right).
\eea
What is important for us is the fact that this charge can be computed
using just the near horizon data \cite{Hanaki:2007mb}, as can be easily
seen from the above expression.

For spinning black holes a
similar definition for the angular momentum exists --- the expression
of the angular momentum does not change in AdS:
\be\label{5ang}
J_{\xi}=  \frac{1}{16 \pi G_5}\int_{H}
\left[* \nabla \xi + 4 * (\xi \cdot A)F + \frac{16}{3 \sqrt{3}}(\xi \cdot A)A \wedge F\right].
\ee
Here, we keep the notations of \cite{Hanaki:2007mb}: $\xi$ is the axial Killing vector
and $\nabla \xi$ is an abbreviation for the two-form
$\nabla_{\mu}\xi_{\nu}dx^{\mu}\wedge dx^{\nu}=d\xi$.

One problem with these five-dimensional definitions is that the Page charges
are not gauge invariant (under large gauge transformations). We note that,
in fact, the Page charge differs from the usual Maxwell charge just by a boundary
term. If this boundary term vanishes (e.g., when the gauge potential decays
sufficiently fast at the boundary) the gauge freedom is removed and so a discussion
in terms of Maxwell charges (computed at the boundary) is equivalent:
$\int_\infty *F = \int_\infty\left(*{}F +\frac{2}{\sqrt{3}} {}A \wedge {}F\right)$.
The supersymmetric spinning black holes of Gutowski and Reall \cite{Gutowski:2004ez}
are written in
such a gauge --- we will present a detailed analysis of these black holes in the
next section.\footnote{From a physical point of view, it is more natural to use
a gauge for which the gauge potential is zero at the horizon. For example, it is
well known that the Euclidean time circle shrinks to a point and to avoid any
problems one shifts $A^H_t=0$ by adding a constant. Therefore, $A^\infty_t=constant$
plays the role of chemical potential of the dual CFT.}

A Kaluza-Klein reduction of Page charges to four dimensions for asymptotically
flat black holes was given \cite{Hanaki:2007mb}. It is straightforward to do a similar analysis
for extremal AdS black holes, but we do not present the details here. Instead, we
use the entropy function to compute the four-dimensional physical charges from
the near horizon data.

We start with the following ansatz for the near horizon geometry and the horizon gauge
field:
\ba
\label{ansatz}
ds^2&=&L_1(-r^2dt^2+r^{-2}dr^2 ) + L_2(d\theta^2 + \sin^2\theta d\psi)
+ L_3^2[d\phi \nn
&& + M_{KK}\cos\theta d\psi + z_1rdt+ A_{KK}(r)dr)]^2
\nonumber \\
A&=&qrdt+M_5\cos\theta d\psi +P d\phi\\
\nonumber
&=&crdt + M_4\cos\theta d\psi + A_4(r)dr+
P[d\phi+ z_1rdt+M_{KK}\cos\theta+A_{KK}(r) dr].
\ea
This ansatz is general enough to describe black branes, non-SUSY
extremal black holes, and spinning {\it supersymmetric} black holes. However,
it is not suitable for extremal non-supersymmetric spinning black holes --- in
this case there is also an angular dependence of some parameters in the near
horizon geometry \cite{Astefanesei:2006dd, nearhorizon}.

Due to the presence of the Chern-Simons term, the action is not gauge invariant. To
apply the entropy function formalism we have to KK reduce to four dimensions
and work with an effective four-dimensional theory --- the details of KK reduction
of the Chern-Simons term are given in Appendix A.

The entropy function is (including the $1/16\pi G_5$ factor  in front of the action)
\bea
E&=& 2 \pi (Q \,c \, R + Z\, z_1\, R- L),\nonumber \\
L&=& \frac{R}{16 \pi G_5} \int d\theta d\psi {\cal L} \nonumber \\
&=&\int_{0}^{\pi} d\theta \int_{0}^{2\pi}d\psi \frac{R \sin \theta}{18} \bigg[9 L_1 L_2 L_3 \bigg(\frac{4 (c+P z_1)^2}{L_1^2}-\frac{4 (P
   M_{KK}+M_4){}^2}{L_2^2}+\frac{4 L_2-L_3^2 M_{KK}^2}{L_2^2}\nonumber \\
   &&+\frac{L_3^2 z_1^2-4
   L_1}{L_1^2}+\frac{24}{M^2}\bigg)-16 \sqrt{3} P \zeta  (P M_{KK} (3 c+2 P z_1)+3 M_4 (2 c+P
   z_1))\bigg]\nonumber \\
   &=& \frac{2 \pi R}{9}\bigg[9 L_1 L_2 L_3 \bigg(\frac{4 \left(c+P z_1\right){}^2}{L_1^2}-\frac{4 \left(P
   M_{KK}+M_4\right){}^2}{L_2^2}+\frac{4 L_2-L_3^2 M_{KK}^2}{L_2^2}\nonumber \\
   &&+\frac{L_3^2 z_1^2-4
   L_1}{L_1^2}+\frac{24}{M^2}\bigg)-16 \sqrt{3} P \zeta  (P M_{KK} (3 c+2 P z_1)+3 M_4 (2 c+P
   z_1))\bigg].
\eea
Here, $R$ is the periodicity of $\phi$ direction, which we would eventually set to identity
in the following computations, thus we have $\int d\phi=R=1$ .

In five dimensions, there are two conserved charges, namely the electric charge
and the angular momentum. In four dimensions, there also are two conserved charges,
namely the $4$-dimensional electric charge, $Q$, and the electric
charge, $Z$, associated to the KK-gauge field. They can be computed in terms of
the near horizon parameters, and we obtain
\ben
\label{4dcharge}
Q&=& \frac{3 c L_2 L_3+P [3 L_2 L_3 z_1-2 \sqrt{3} L_1 (P M_{KK}+2 M_4)]}{3 G_5
   L_1}, \\
\nonumber
Z&=&\frac{36 c L_2 L_3 P-8 \sqrt{3} L_1 P^2 (2 P M_{KK}+3 M_4)+9 L_2 L_3 z_1 (L_3^2+4
   P^2)}{36 G_5 L_1}.
\een
These charges are the gauge invariant Maxwell charges that characterize the
$4$-dimensional solution.

Let us now compute the $5$-dimensional Page charges given in (\ref{page}) and (\ref{5ang}).
These charges are not gauge
invariant, but let us fix the gauge so that the $5$-dimensional gauge field
matches (\ref{ansatz}):
\ben
A_5&=& A_4 + P (d \phi + A_{kk}), \nonumber \\
F^5_{ab}&=&F^4_{ab} + P F^{kk}_{ab} +\partial_{a}P A^{kk}_{b}- \partial_{b}P A^{kk}_{a}.
\een
where $A_4$ and $A^{kk}$ can be read off from (\ref{ansatz}) and $a,b,..$ denote
the four-dimensional indices.\footnote{For simplicity we consider  $\int d\phi=1$ in the
ret of this section.} We obtain the
following expression for the electric charge:
\ben
Q_{5d}&=&- \frac{1}{4 \pi G_5} \int_{H}d \theta d \psi
\left[\frac{L_2 L_3}{L1}(c+P z_1)\sin\theta - \frac{2}{\sqrt{3}} 2 P
F^5_{\psi \theta}\right] \nonumber \\
&=&- \frac{1}{4 \pi G_5} 4 \pi \left[\frac{L_2 L_3}{L1}(c+P z_1) -
\frac{2}{\sqrt{3}} 2( P M_4 +\frac{1}{2} P^2 M_{kk})\right]\\
\nonumber
&=&- \frac{1}{ G_5}  \left[\frac{L_2 L_3}{L1}(c+P z_1) -
\frac{2}{\sqrt{3}} (2 P M_4 + P^2 M_{kk})\right].
\een
We emphasize that in the Chern-Simons term we have done the integration by parts.

A similar computation can be done for the angular momentum, and we obtain
\ben
J_{\xi}&=& \frac{1}{4 \pi G_5}\int_{H}\bigg[\frac{1}{4}* \nabla \xi + * (\xi \cdot A)F
+ \frac{4}{3 \sqrt{3}}(\xi \cdot A)A \wedge F\bigg]\nn
 &=& -\frac{1}{4 \pi G_5}\int_{H} d \theta d \psi \bigg[-\frac{1}{4}(*
\nabla \xi)_{\psi\theta}
+ P  \frac{L_2 L_3}{L_1}(c+P z_1)\sin\theta - \frac{4}{3
\sqrt{3}}\frac{3}{2}P^2 F^5_{\psi \theta}\bigg]
\nn
&=&-\frac{1}{4 G_5}\bigg[\frac{L_2 L_3^3 z_1}{L_1}  +
4 P  \frac{L_2 L_3}{L_1}(c+P z_1)- 4\frac{2}{3 \sqrt{3}}P^2(3 M_4+2 P M_{kk})\bigg].
\een
Now, one can compare these results with (\ref{4dcharge}) and so we indeed
obtain a match between four-dimensional and five-dimensional charges:
\be
Q_{5d}= - Q_{4d}.
\ee
Thus, we see that the $4$-dimensional charges of extremal asymptotically AdS black
holes are the dimensional reduced $5$-dimensional Page charges for a fixed gauge.

Since in four dimensions there is no Chern-Simons term, the physical charges are
the Maxwell charges. These charges are conserved and localized. The $A_z$ component
of the five-dimensional gauge potential becomes a scalar from a four-dimensional
point of view and so the four-dimensional Maxwell charges are shifted because of
this term. Since the Page charges are not gauge invariant, they are not the same
with the gauge invariant four-dimensional Maxwell charges. However, one can compare
them once the gauge is fixed. From a five-dimensional point of view, due to the
Chern-Simons term, there is hair (degrees of freedom outside the horizon) which
contributes to the charges. In four dimensions, the interpretation is completely
different: there is no hair, but there exists an extra scalar which is fixed (due
to the attractor mechanism) at the horizon.

\section{Examples}
In this section, we apply the previous general results
in a few concrete examples. We compute the physical charges and find
the near horizon geometry of baryonic and electromagnetic black
branes, and also for supersymmetric spinning black holes.

\subsection{Baryonic black branes}
In \cite{Herzog:2009gd}, a new type of black 3-brane charged under a baryonic
$U(1)_B$ was obtained and a numerical analysis in the limit $T\rightarrow 0$
was presented. However, when the horizon is located deep inside the
near $AdS_2$ region (small temperature) it is difficult to study
the system numerically.

In what follows, we use the results of the previous
section to {\it analytically} obtain the near horizon data of the
baryonic black branes at zero temperature. Our analysis confirms the
results of \cite{Herzog:2009gd}. However, we would like to point
out that there is an important problem with this solution. That is,
the radius of $AdS_2$ is blowing up.

We are interested in the following effective action, which is a consistent
truncation of type $IIB$ SUGRA:
\be
S=\int d^5x \sqrt{g}\bigg[ R - \frac{1}{4}e^{2 \eta - \frac{4}{3}\chi}
F_{\mu \nu}F^{\mu \nu}-5 \partial_{mu}\eta \partial^{mu}\eta -
\frac{10}{3}\partial_{mu}\chi \partial^{mu}\chi- V(\eta,\chi)\bigg],
\ee
where the potential of the two neutral scalars is \cite{Herzog:2009gd}
\be
V(\eta,\chi)=\frac{8}{L^2}e^{-\frac{20}{3}\chi}+\frac{4}{L^2}
e^{-\frac{8}{3}\chi}(e^{-6 \eta}-6 e^{-\eta})
\ee
and $L^4 = 4\pi g_s N (\alpha')^2 \frac{27}{16}$ is obtained from the quantization
of the five-form flux.

We are going to study extremal solutions with the following near horizon geometry:
\ben
ds^2= L_1 (-r^2 dt^2+\frac{dr^2}{r^2})+ L_2 (dx^2+dy^2+dz^2) \nonumber \\
A= q r dt, \,\,\ \eta= \eta_0, \,\, \chi=\chi_0.
\een
Here, A is the $U(1)_B$ gauge potential; $\eta_0$ and $\chi_0$ are the
horizon values of the neutral scalars. Since we consider a static solution
in a theory without Chern-Simons term, there is no fibration in the near
horizon geometry. We will see in the next section that, for electromagnetic
branes, there is a non-trivial fibration of $AdS_2$ in the near horizon
geometry.

As usual, the entropy function is $E = 2 \pi ( q Q - {\cal L})$, where
${\cal L}$ is the Lagrangian density at the horizon:
\begin{equation}
\nonumber \\
{\cal L} = \int_H dx dy dz \sqrt{g}\bigg[ R -
\frac{1}{4}e^{2 \eta - \frac{4}{3}\chi}F_{\mu \nu}F^{\mu \nu}
-5 \partial_{mu}\eta \partial^{mu}\eta -
\frac{10}{3}\partial_{mu}\chi \partial^{mu}\chi- V(\eta,\chi)\bigg].
\end{equation}
The equations of motion in the near horizon geometry (attractor equations) are
\ben
L_2^{3/2} e^{-6 \eta_0 -\frac{20 \chi_0 }{3}} \left[\frac{L^2 q^2 e^{8 \eta_0 +\frac{16 \chi_0 }{3}}}{L_1^2}-8 \left(6 e^{5 \eta_0
   }-1\right) e^{4 \chi_0 }+16 e^{6 \eta_0 }\right]&=&0, \nonumber\\
L_1 \sqrt{L_2} \left[\frac{1}{2} e^{-20 \chi_0 /3} \left(\frac{q^2 e^{2 \eta_0 +\frac{16 \chi_0 }{3}}}{L_1^2}-\frac{8 \left(\left(1-6
   e^{5 \eta_0 }\right) e^{4 \chi_0 -6 \eta_0 }+2\right)}{L^2}\right)-\frac{2}{L_1}\right]&=&0 ,\nonumber \\
L_2^{3/2} e^{-6 \eta_0 -\frac{8 \chi_0 }{3}} \left[\frac{L^2 q^2 e^{8 \eta_0 +\frac{4 \chi_0 }{3}}}{L_1}-24 L_1 \left(e^{5 \eta_0
   }-1\right)\right]&=&0, \nonumber \\
 L_2^{3/2} e^{-6 \eta_0 -\frac{20 \chi_0 }{3}} \left[\frac{L^2 q^2 e^{8 \eta_0 +\frac{16 \chi_0 }{3}}}{L_1}-16 L_1 \left(-6 e^{5 \eta_0
   +4 \chi_0 }+5 e^{6 \eta_0 }+e^{4 \chi_0 }\right)\right]&=& 0, \nonumber \\
Q-\frac{L_2^{3/2} q e^{2 \eta_0 -\frac{4 \chi_0 }{3}}}{L_1}&=&0. \nonumber \\
\een
The last equation relates the gauge potential to the physical (baryonic) charge of
the system:
\be
q=\frac{L_1 Q e^{\frac{4 \chi_0 }{3}-2 \eta_0 }}{L_2^{3/2}}.
\ee
Next, we rewrite the other equations in terms of $Q$
and solve for the near horizon parameters.

We are interested in regular solutions and so we assume that the radii of $AdS_2$ and
$R^3$ are non-zero quantities (so that we can multiply and divide by $L_1, L_2$).
First, we consider combinations of the first two equations and we obtain
\ben
\frac{L^2}{L_1}+4 e^{-6 \eta_0 -\frac{20 \chi_0 }{3}} \left(-6 e^{5 \eta_0 +4 \chi_0 }+2 e^{6 \eta_0 }+e^{4 \chi_0 }\right)&=&0 ,\nonumber \\
Q^2 e^{\frac{4 \chi_0 }{3}-2 \eta_0 }-\frac{2 L_2^3}{L_1}&=&0.
\een
In this way we get the radii of $AdS_2$ and $R^3$ in terms of the near
horizon values of the scalars:
\ben
L_1&=&-\frac{L^2 e^{6 \eta_0 +\frac{20 \chi_0 }{3}}}{4 \left(-6 e^{5 \eta_0 +4 \chi_0 }+2 e^{6 \eta_0 }+e^{4 \chi_0 }\right)},\nonumber \\
L_2&=&-\frac{L^{2/3} Q^{2/3} e^{\frac{4}{3} (\eta_0 +2 \chi_0 )}}{2 \sqrt[3]{-6 e^{5 \eta_0 +4 \chi_0 }+2 e^{6 \eta_0 }+e^{4 \chi_0 }}}.
\een
We replace these expressions in the remaining equations of motion and
solve for
the horizon values of the scalars:
\be
\eta_0 - 4 \chi_0 = \log {{3 \over 2}}\, , \,\,\,\,\ e^{- \eta_0}=0.
\ee
We reobtain the results of \cite{Herzog:2009gd}, namely the scalars are blowing up at the
horizon for the extremal solutions. This is already a sign that the extremal
solution may be problematic.

The advantage of the entropy function is that we can analytically compute the
other horizon parameters. First, we observe that even if the horizon values of
the scalars diverge, their combination, which enters in the entropy function,
is finite. This is why we obtain a finite entropy density
\be
S= \frac{2}{3}\pi Q.
\ee
which is an indication that the radius of $R^3$, $L_2$, is finite --- indeed, from
the above analysis, we
obtain $L_2 = ({L Q / 6})^{2/3}$ . However, the radius
of $AdS_2$, $L_1$, is blowing up. This should not come as a surprise, since the
horizon values of the moduli are not finite. Therefore we see that $AdS_2$
isometry does not exists in the near horizon geometry. Following \cite{nearhorizon},
we would like to comment that for this baryonic black brane system even a flat near-horizon
extremal geometry is not possible. It will be interesting to check if
in the presence of stringy corrections there exist regular solutions.

\subsection{Stationary black holes}

\subsubsection{Electromagnetic black branes}
A study of black brane
solutions of Einstein-Maxwell AdS gravity with a gauge Chern-Simons
term can be found in \cite{D'Hoker:2009mm, Astefanesei:2010dk}.
These solutions are important for understanding how the magnetic fields
and Chern-Simons term affect the hydrodynamic properties of
the dual field theories.

In \cite{Astefanesei:2010dk}, we have used the
entropy function to carefully study the near horizon geometry of
these black branes at zero temperature and discuss different branches
with finite area horizons. We have presented analytic expressions
for the entropy density that support the numerical analysis of
D'Hoker and Kraus (though, we find a larger class of near
horizon geometries). That is, there is a critical value of the magnetic
field for which the entropy vanishes.

In this work we are interested to understand the role of the Chern-Simons
term for the definition of the physical charges. We know that, due to the
Chern-Simons term, the near horizon geometry contains, in fact, a non-trivial
fibration of $AdS_2$ even if these solutions are not spinning. We start with the
near horizon geometry and use the entropy function formalism to compute
the four-dimensional charges. Then, we are going to identify the contribution
from the Chern-Simons term.

We consider
the $z$-direction compactified on a circle of
radius $\beta$ and a general KK ansatz
\ben
g_{\alpha\beta}dx^{\alpha}dx^{\beta}&=&G_{ab}dx^adx^b+G_{AB}(dy^A+\bar
{A}_a^Adx^a)(dy^B+\bar{A}_a^Bdx^a),\nn
A^{(5)}&=&A_\mu^{(5)}dx^\mu = A_a^{(4)}dx^a + C_B(x^a) (dy^B + \bar{A}^B_a dx^a) ,
\een
where ${a,b}$ are $4$D indices and ${A,B}$ are compact indices
($z$ in our case). With this notation we have splitted the
coordinates as $x^\mu=(x^a, y^A)$ and so $A_\mu^{(5)}$ is the
$5$-dimensional gauge potential, $A_a^{(4)}$ is the $4$-dimensional
gauge potential, and $\bar{A}^B_a$ is the KK gauge potential.

After KK reduction, the metric, gauge field, and
relations between the $5$-dimensional gauge potential, KK gauge potential, and $4$-dimensional
gauge potential are
\ben
\label{AnsatZ}
ds^2&=& L (-\frac{dt^2}{r^2}+r^2dr^2) + \upsilon_1 (dx^2+  dy^2) +
\upsilon_2(dz + z_1 r dt)^2, \nn
A_{\mu}^{(5)}dx^{\mu} &=& \vartheta r dt - \frac{B}{2} y dx +
\frac{B}{2} x dy - p_1 (dz+z_1 r dt),\nn
F_{rt}^{(5)}&=& q = \vartheta - p_1 z_1\, , \,\,\, F^{(5)}_{xy}=F^{(4)}_{xy}=B\, , \,\,\,  F_{rt}^{(4)} = \vartheta \, , \,\,\,
\bar{F}_{rt}^{z} = z_1.
\een
The on-shell action and entropy function are
\ben
S&=& \frac{{\cal A}_{xy} \beta} {16 \pi G_5} \left[L \sqrt{\upsilon_1^2 \upsilon_2}
\left(-\frac{2 B^2}{\upsilon_1^2}+\frac{2 (\vartheta-p_1 z_1)^2}{L^2}
+\frac{\upsilon_2 z_1^2-4 L}{2 L^2}+12\right)+8 B p_1 \zeta
\lb\frac{p_1 z_1}{2} -\vartheta\rb\right], \nn
{\cal E} &=& 2 \pi \beta {\cal A}_{xy} \lb Q\  \vartheta + \Theta\  z_1- {S\over \beta {\cal A}_{xy}}\rb ,
\een
where $Q$ is the $4$-dimensional physical charge, $\Theta$ is the physical
charge associated to $KK$ gauge field, and ${\cal A}_{xy}=\int dx dy$.

The relevant equations are (more details can be found in \cite{Astefanesei:2010dk})
\ben
 \frac{2 \beta \pi}{L}\left[L\Theta +\frac{1}{16 \pi G_5}(\upsilon_1
\sqrt{\upsilon_2} (4 p_1 q-\upsilon_2 z_1)-4 B p_1^2 \zeta L
)\right]&=&0,\nn
2 \pi  \beta \left[\frac{1}{4 \pi G_5}\lb B p_1 \zeta -\frac{ \upsilon_1 \sqrt{\upsilon_2} q}{L}\rb+Q\right] &=& 0.
\een

From these equations it is clear that the four-dimensional charges have a contribution
that contains the Chern-Simons coupling, $\zeta$. These charges are Maxwell charges and
the extra contribution is due to the new scalar associated to the compact direction (the
modulus $p_1$).

To compare with the five-dimensional Page charge, we should compute the
expression (\ref{page}) (at the horizon) by using the near horizon geometry
(\ref{AnsatZ}). This can be easily done and, as expected, the charges match.
As we have already explained, there is a non-trivial fibration of $AdS_2$ in the near
horizon geometry. The momentum along the $z$-circle becomes, after KK reduction,
the $4$-dimensional electric charge, $\Theta$. The way the near horizon geometry is
fibered will correspond to the magnetic charge. However, since the magnetic charge
is topological it can not be fixed by the entropy function.

\subsubsection{Susy spinning AdS black holes}
In flat space \cite{Reall:2002bh}, the geometry of the event horizon of
any supersymmetric black hole of minimal $5$-dimensional supergravity must
be $T^3$, $S^1 \times S^2$, or a quotient of a homogeneously squashed $S^3$.
However, there is no general classification of the near horizon
geometries of SUSY black holes in AdS spacetime.

In AdS space \cite{Gutowski:2004ez},  Gutowski and Reall
found an interesting solution that is asymptotically AdS and does not
have an $AdS_3$ component in the near-horizon geometry. In the ungauged
theory the near-horizon geometry of a BPS black hole is maximally supersymmetric. In
the gauged supergravity this is not true because the only maximally supersymmetric
solution is $AdS_5$.

We use the coordinates that make the $AdS_2$ part of the near
horizon geometry manifest \cite{Astefanesei:2009sh}:
\ba
\label{GR}
ds^2 &=&\frac{1}{\Delta^2 + 9 \ell^{-2}}(-r^2d\tau^2+r^{-2}dr^2) +\frac{1}{\Delta^2 - 3 \ell^{-2}}(d\theta^2 +
\sin^2 \theta d\psi^2)
\nn
&+&\left(\frac{\Delta}{\Delta^2 - 3 \ell^{-2}}\right)^2 \left[d\phi + \cos \theta d \psi
 - \frac{3r \alpha}{\ell\Delta}\frac{\Delta^2 - 3 \ell^{-2}}{\Delta^2 + 9 \ell^{-2}}(d\tau+\frac{dr}{r^2}) \right]^2
\ea
%
and the gauge potential
\ba
A=\frac{\sqrt{3}}{2(\Delta^2+9l^{-2})} \Delta r d\tau
- \frac{\sqrt{3} \alpha \cos
\theta}{ 2 \ell (  \Delta^2 - 3 \ell^{-2} ) }d\psi
\ea
where, $\Delta > \frac{\sqrt{3}}{\ell}$ and $\alpha= \pm 1$.

Since we have an exact solution, our goal is not to use the entropy
function for finding the near horizon geometry. We want to use the
entropy function to compute the corresponding $4$-dimensional (physical)
Maxwell charges from the attractor equations. First, we do KK reduction
to obtain a gauge invariant effective action. Then, we check that the
near horizon geometry of Gutowski-Reall black hole is a solution. Finally,
we solve the gauge fields' equations of motion to obtain the physical
charges from the near horizon geometry.

It is straightforward to check that the near horizon geometry and gauge
field (\ref{GR}, \ref{KKGR}) are solutions of the attractor equations
obtained from the general ansatz (\ref{ansatz}).

Before presenting the charges, let us make a couple of observations. First,
by using the equations of motion for $L_1$ and $L_2$ we obtain that
$L_1 \left(12 L_2 + l^2\right)-L_2 l^2=0$. This is expected \cite{Astefanesei:2007vh}, since
in $AdS$ spacetime the radii of $AdS_2$ and $S^3$ in the near horizon geometry
are not equal --- they are related by the cosmological constant (more generally,
by the potential of the moduli). The second observation is
related to the entropy. That is, the parameter $P$ can not be fixed by the equations
of motion in the near horizon limit. The reason is that in the work \cite{Gutowski:2004ez} the
convention for the Chern-Simons coefficient is $\zeta=1$ and the equation
of $P$ is trivially satisfied in this way. One may think that the fact that
$P$ is not fixed causes problems for the entropy --- $P$ appears in the entropy
function and is not fixed. However, a concrete computation of the entropy
reveals that $P$ cancels trivially and does not appear in the final expression
of the entropy:
\be
S=\frac{16 \pi^2 R l^4 \Delta }{(l^2 \Delta^2-3)^2}.
\ee
where $R$ is the periodicity of $\phi$.

However, the expressions of the $4$-dimensional charges depend on $P$. Since
we have an exact solution we read off $P$ from the near horizon geometry and
get:
\be
P= \frac{\sqrt{3} M \alpha }{6-2 M^2 \Delta ^2}.
\ee
We also obtain the following $4$-dimensional charges:
\bea
Q = \frac{\sqrt{3}  M^2 \left(M^2 \Delta ^2-1\right)}{2 G_5\left(M^2 \Delta ^2-3\right)^2}\,\,\, , \,\,\,\,\,\,\,\,\,\,\,\,\,\,\,
Z = \frac{ M^3 \alpha  \left(1-3 M^2 \Delta ^2\right)}{2 G_5\left(M^2 \Delta ^2-3\right)^3}.
\eea
These charges should be used when comparing the thermodynamical and statistical
entropies.

For completeness, let us also present the expressions of the five dimensional
charges: \footnote{Compared to \cite{Gutowski:2004ez},  our definitions for the five-dimensional
charges  should be divided by $4 \pi$. The reason is that we have considered
the periodicity of $\phi$, $R=1$,  in the previous general discussion in Section $2$.}

\ben
Q_{5d} =\frac{\sqrt{3} r_0^2 \left(2 M^2+r_0^2\right)}{16 G_5 M^2} \,\,\, , \,\,\,\,\,\,\,\,\,\,\,\,\,\,
J = \frac{ r_0^4 \alpha  \left(3 M^2+2 r_0^2\right)}{32 G_5 M^3}
\een
where $r_0$ is the horizon radius.
We notice that while $Q_{5d}$ is independent of the parameter $\alpha (= \pm 1)$, the angular momentum
$J$ does depend on it. Now, using the relation between the horizon radius $r_0$ and near horizon parameter
$\Delta$, $r_0=2 M \sqrt{\frac{1}{M^2 \Delta ^2-3}}$, we get the expected match between the
four-dimensional and five-dimensional charges. We would like to point out that the solution
is written in a gauge for which the Page charge matches the electric five dimensional Maxwell charge.

\section{Discussion}
\label{discuss}
We have used the entropy function formalism to compute the physical charges of a
general class of extremal AdS black holes. These new results combined with the
method of \cite{Hanaki:2007mb} prove that, indeed, the physical charges in the
presence of the gauge Chern-Simons term are the Page charges. This closes a gap
in computing the physical charges from the near horizon data for extremal AdS black
holes.

The focus of this paper was on the gauge Chern-Simons term and we did not consider
the gravitational Chern-Simons term, which is a four derivative metric term. One expects
that again the charges will be shifted in a similar manner. Indeed, the
shift in four-dimensional and five-dimensional charges found in
\cite{Castro:2008ys,dewit-katmadas} is proportional to the above mentioned
term. Their analysis was done for asymptotically flat black holes.

We then checked our results in some explicit examples. In particular, we have
obtained analytic results for the near horizon geometry of baryonic branes at
zero temperature. Since these are extremal solutions, the near horizon geometry
contains an $AdS_2$ spacetime. We have found that the radius of $AdS_2$ is
blowing up, and from this point of view the solution is problematic. However,
it still remains possible that a general set of stringy corrections would allow
regular solutions in this case.

An analysis of the physics of AdS black holes using the near horizon data only is
also important for computing the shear viscosity to entropy density ratio due to
the AdS/CFT duality (for other transport coefficients there are, in general, non-trivial
flows and so the near horizon geometry is not sufficient). In the presence of the
gauge Chern-Simons term, one has to work with the right physical charges in five
dimensions. These charges are the Page charges (not the usual Maxwell charges) and
we have proven that they consistently match the physical charges obtained from
the entropy function. Due to the Chern-Simons term, there is a non-trivial fibration
of $AdS_2$ in the near horizon geometry.

When there is a compact direction, after KK reduction, we can compare the five dimensional
Page charges with the Maxwell charges in four dimensions. We have found that, indeed, they
match as suggested by \cite{Hanaki:2007mb}. However, our interpretation is slightly different.
The reason is that the entropy function formalism provides enough information to
exactly understand the meaning of the extra contribution from the gauge Chern-Simons
term. Since we have presented in detail these arguments in Section 2, here we would like
just to point out that the interpretation is completely different in four dimensions. That is, there is no hair, but there exists an extra scalar that is fixed (due to the attractor
mechanism) at the horizon. The extra contributions are due to this modulus.

Finally, we have also investigated SUSY spinning black holes in AdS. We have computed the
charges for an exact solution and found again the expected agreement. This solution has a
spherical horizon and one can wonder how our results will apply to spinning black holes
with toroidal horizon geometry. If one can find such a solution, it will be also interesting
to compute the transport coefficients of the dual plasma. Unfortunately, this solution
does not exist.\footnote{We would like to thank Harvey Reall for clarifications on this
point.} This observation was made in \cite{Gutowski:2004ez}, where it was proven that the
corresponding supersymmetric near-horizon solution joins, in fact, to a solution which is
asymptotically a plane-wave geometry and not AdS as we need. In Appendix B, we use the method
of \cite{Gutowski:2004ez} to show that there can not exist an asymptotically AdS solution, which
can join up with this near horizon geometry and preserves $\frac{1}{4}$ SUSY
(which is minimal in this case). Though, as a final remark, we would like to point out that
there may exist non-supersymmetric extremal spinning AdS black holes with toroidal horizon geometry.

\section*{Acknowledgements}
We would like to thank Guillaume Bossard, Yuji Tachikawa, Stefan Theisen,
Stefan Vandoren, Oscar Varela and Bernard de Wit for interesting discussions
and Ashoke Sen for valuable discussions and comments on an earlier draft of this paper.
DA would also like to thank PUCV Chile, Universidad Catolica de Chile,
and CECS Valdivia for the hospitality during part of this work and
the audience at all these places for their positive feedback. Work of NB
is supported by NWO Veni grant, the Netherlands.

\vspace{1cm}

\appendix

\noindent
{\Large \bf Appendix}

\section{KK reduction of Chern-Simons term}
We would like to use entropy function formalism to compute the physical
charges. Since the $5$-dimensional action is not gauge invariant,
we should KK reduce to four dimensions. In this section we explicitly
show the KK reduction of the Chern-Simons term.

First, we observe that the most general gauge potential compatible
with the symmetries of the metric (\ref{GR}) is
\ba
\label{KKGR}
A=\frac{\sqrt{3}}{2(\Delta^2+9l^{-2})} \Delta r d\tau
- \frac{\sqrt{3} \cos
\theta}{ 2 \ell (  \Delta^2 - 3 \ell^{-2} ) }d\psi + b\left[d\phi+   \cos \theta d \psi
 - \frac{3r}{\ell\Delta}\frac{\Delta^2 - 3 \ell^{-2}}{\Delta^2 + 9 \ell^{-2}}(d\tau+\frac{dr}{r^2})\right].
\ea
Unlike the other cases (e.g., SUSY black holes in flat space) studied
in the literature, the KK gauge potential has also a magnetic
part. Since $A^{KK}_r$ is a function of the radial coordinate
only, it does not play any role in our analysis (contributions
of the form $\epsilon^{\mu\nu\alpha rr}\partial_r A_r$ are
zero in the Chern-Simons term).

Let us consider a generic ansatz for the gauge potential:
\be
A = A_{\mu}dx^{\mu} = B_{a}dx^{a} + b(x^a)[ d\phi + A^{KK}_{a}dx^{a}]
\ee
where by ${\mu,\nu,..}$ we denote the $5$-dimensional indices
and by ${a,b,..}$ the $4$-dimensional ones. The Chern-Simons term becomes
\be
\epsilon^{\alpha\beta\gamma\delta\mu}A_\alpha F_{\beta\gamma}F_{\delta\mu}=
3\epsilon^{\beta\gamma\delta\mu}A_\phi F_{\beta\gamma}F_{\delta\mu}
\ee
and its contribution to the action is
\ben
CS &=& {\zeta \over 16\pi G_5} R \int d^4 x \epsilon^{abcd} A_{\phi}
F_{ab}F_{cd}, \nn
F_{ab}&=&\partial_a A_b - \partial_b A_a.
\een
where $R$ is the periodicity of $\phi$ (KK compact direction).

To rewrite this part in a suitable form, one should also integrate
by parts terms of the form $b\partial_{\mu}b$ or $b^2\partial_{\mu}b$.
The final result (in a covariant frame) is
\be
CS = {4 R \zeta \over 16\pi G_5}  \int d^4 x
\epsilon^{abcd}[{1 \over 3}b^3\partial_a A^{KK}_b \partial_c A^{KK}_d +
b \partial_a B_b \partial_c B_d
+b^2 \partial_a A^{KK}_b \partial_c B_d].
\ee

To illustrate this method, let us consider a simple example when there is no
KK magnetic charge:
\begin{equation}
A=erd\tau + p\cos\theta d\psi + b[\bar{e}rd\tau + A_r(r)dr].
\end{equation}
The relevant part of the Chern-Simons term is
\ba
bF_{rt}F_{\theta \psi}=b\,p\,\partial_r A_\tau = p [ be + \bar{e}\,b \partial_r (br)]=
p [ be + \bar{e}(b^2+br\partial_rb)] = p [be + \bar{e}(b^2+\frac{r}{2}\partial_rb^2)].
\ea
The last term is then $b^2 -b^2/2 + \partial_r(rb^2)/2$ and we see that after we get rid of
the total derivative, we obtain $pb(e+b\bar{e}/2)$.

\section{Toroidal Spinning Black Brane}

In this appendix, we will discuss another possible near-horizon geometry that arises in 
five-dimensional minimal gauged supergravity, namely $AdS_2 \times T^2 \times S^1$. This 
corresponds to $\Delta=\frac{\sqrt{3}}{\ell}$ \cite{Gutowski:2004ez}:
\ba
\label{eqn:sol1}
 ds^2 &=& - \frac{3 r^2}{\ell^2} du^2 + 2 du dr - \frac{6r}{\ell}du \left(dz
+ \frac{\sqrt{3}}{2 \ell} (y dx - x dy) \right) \nn &+& \left(dz +
\frac{\sqrt{3}}{2 \ell} (y dx - x dy) \right)^2 + dx^2 + dy^2 ,\nn
F &=& -\frac{3}{2\ell}du \wedge dr + \frac{\sqrt{3}}{2\ell} dx \wedge dy.
\ea

This metric can be rewritten as
\be
ds^2= -\frac{12 r^2}{l^2}du^2+2 du dr+\bigg((dz +
\frac{\sqrt{3}}{2 \ell} (y dx - x dy))- \frac{3 r}{l}du \bigg)^2 + dx^2+dy^2
\ee
and with the following coordinate transformation
\be
\tau= \frac{12}{l^2}u+\frac{1}{r}.
\ee
we explicitly obtain the $AdS_2$ part:
\be
ds^2=\frac{l^2}{12}\bigg(-r^2 d\tau^2 +\frac{dr^2}{r^2}\bigg)+ \bigg((dz +
\frac{\sqrt{3}}{2 \ell} (y dx - x dy))- \frac{l r}{4}(d \tau+\frac{dr}{r^2}) \bigg)^2 + dx^2+dy^2.
\ee
The gauge field strength becomes
\be
F= - \frac{3 l}{24} d\tau \wedge dr +\frac{\sqrt{3}}{2\ell} dx \wedge dy.
\ee

In \cite{Gutowski:2004ez}, the authors have shown that this supersymmetric 
near horizon geometry corresponds to a solution, which is a plane-wave 
geometry. In this appendix we will show that there is no asymptotically AdS 
with this near horizon geometry such that it preserves $\frac{1}{4}$ SUSY. To 
show this, we will proceed as in \cite{Gutowski:2004ez}. We will try to construct 
a base space solution that can give us a global AdS black brane geometry. It was 
proved in \cite{Gauntlett:2002nw} that a SUSY solution of this theory can be 
completely parameterized in terms of a scalar $f$, a real vector $V^\alpha$, and 
three real two-form fields $X^i$. These quantities satisfy the following algebraic 
relations:
\ben
V^{\alpha}V_{\alpha}&=&-f^2, \nn
X^i \wedge X^j &=& -2 \delta_{ij} f \star V,\nn
\iota_V X^i &=& 0, \nn
\iota_V \star X^i &=& = - f X^i,\nn
(X^i)_{\gamma\alpha}(X^J)_{\beta}^{\gamma} &=& \delta_{ij} (f^2
\eta_{\alpha\beta}+V_{\alpha}V_{\beta}) -
\epsilon_{ijk}f(X^k)_{\alpha\beta}.
\een
They also satisfy the following differential relations:
\ben\label{fco}
df &=& - \frac{2}{\sqrt3}\iota_V F,\nn
D_{(\alpha}V_{\beta)} &=&0, \nn
dV &=& - \frac{4}{\sqrt3} f F - \frac{2}{\sqrt3}\star (F\wedge V) -
2l^{-1} X^1, \nn
dX^i &=& \frac1l \epsilon_{1ij}\ltb2\sqrt3 A\wedge X^j + 3 \star X^j
\rtb .
\een

We would like to construct a solution for which $V$ is globally defined 
and it can always be made timelike in a coordinate patch. In such a coordinate patch,
the metric can be written in a nice form 
\be
ds_5^2= -f^2(dt +\omega)^2 + f^{-1}ds_4^2,
\ee
and without the loss of generality the scalar $f$ can be chosen to be positive
and $V=\frac{\partial}{\partial t}$. Here, $ds^2_4$ is the metric of the base 
space ${\cal B}$ and $\omega$ is a one-form on ${\cal B}$. Imposing the supersymmetric 
constraints listed above and integrability, one finds that  ${\cal B}$ has to be a Kahler 
manifold with $X^1$ the anti-selfdual Kahler form. Also the scalar 
$f= -\frac{24}{R}$, where $R$ is the Ricci scalar of the base space. Now let us first write 
down the Kahler base space metric and the corresponding Kahler for the near horizon 
geometry (\ref{eqn:sol1})
\ben\label{bsnh}
ds_4^2&=& dr^2 + r^2(\sigma_1^2+\sigma_2^2)+4 r^2 \sigma_3^2 \nn
X^1&=& \sqrt{3}d[r^2 \sigma_3],
\een
where we have defined three one-forms
\be
\sigma_1=\sqrt{3} dx,\,\,\ \sigma_2=\sqrt{3} dx,\,\,\  \sigma_3= \sqrt{3}(dz+\frac{\sqrt{3}}{2}(y dx- x dy)).
\ee
Next, we write the base space and the two-form corresponding to the asymptotically AdS space 
that joins up with the near horizon metric (\ref{eqn:sol1}) as
\ben
ds_4^2&=& dr^2 + A^2(r)(\sigma_1^2+\sigma_2^2)+4 B(r)^2 \sigma_3^2 \nn
X^1&=& d[C(r) \sigma_3],
\een
where $A(r), B(r),C(r)$ are arbitrary functions of the radial coordinate r 
so that, near the horizon, they are given by (\ref{bsnh}).

Since the AdS spacetime is maximally symmetric, it requires the gauge field 
strength to vanish and due to first equation of (\ref{fco}) the scalar $f$ is constant. Without 
loss of generality, we can choose it to be identity. Now, the anti-selfduality 
of $X^1$ gives
\be
C(r)=A(r)^2,\,\,\,\,\,\ B(r)=2 A'(r) A(r) .
\ee
Thus effectively we have to find only one function $A(r)$ to fix the geometry. Moreover, 
since the scalar function $f$ is a constant in this case and has been set to one, the Ricci 
scalar $R$ of the base space metric should be a constant, namely $R=-24$. Since the Ricci 
scalar is
\be
R=- \frac{2(4 A'(r)^2+7 A(r)A''(r))}{A(r)}-\frac{2 A(r)^3}{A'(r)},
\ee
a solution is $A(r)=e^{r}$.

It was shown in \cite{Gutowski:2004ez} that another necessary condition for the geometry to 
be asymptotically AdS is 
\be
R_{ijpq}= -2 X^1_{ij}X^1_{pq} + (X^1_{qi}X^1_{pj}-X^1_{pi}X^1_{qj})-(\delta_{pi}\delta_{qj}-\delta_{qi}\delta_{pj}),
\ee
where $(i,j,p,q)$ are tangent space indices. This condition gets
satisfied automatically with the above choice of the metric function
$A(r)=e^{r}$ . The SUSY geometry can then
be written as
\be
ds^2 = -\lb dt + 2 e^{2r} \sigma_3 \rb^2 + 3e^{2r}\lb dx^2+dy^2
+4e^{2r}\lb dz+ \frac{\sqrt{3}}{2} (yds-xdy)\rb^2 \rb + dr^2.
\ee
With a coordinate transformation $e^r = \rho$
we can easily see that it is a PP-wave geometry:
\ben
ds^2 &=& - \lb dt + 2 \rho^2 \sqrt{3} \lb dz+ \frac{\sqrt{3}}{2}
(yds-xdy)\rb\rb^2 + \frac{d\rho^2}{\rho^2} \nn
&& + 3 \rho^2
(dx^2 + dy^2) + 12 \rho^4 \lb dz+ \frac{\sqrt{3}}{2}
(yds-xdy)\rb^.
\een

Therefore, we conclude that there is no asymptotically AdS
supersymmetric spinning black brane solution with (\ref{eqn:sol1}) as
the near-horizon geometry.

\end{document}